\documentclass{aa}
\usepackage{graphicx}
\usepackage{natbib}
\bibpunct{(}{)}{;}{a}{}{,}   

\newcommand{\eqb}{\begin{eqnarray}}
\newcommand{\eqe}{\end{eqnarray}}

\newcommand{\nelec}{n_{\rm e}}

\newcommand{\tcross}{t_{\rm cr}}
\newcommand{\gammamax}{\gamma_{\rm max}}
\newcommand{\gammamin}{\gamma_{\rm min}}

\newcommand{\qcal}{\cal{Q}}
\newcommand{\nph}{n_{\rm \gamma}}
\newcommand{\eB}{\epsilon_{\rm B}}
\newcommand{\ee}{\epsilon_{\rm e}}

\newcommand{\ke}{k_{\rm e}}
\newcommand{\Rdec}{R_{\rm dec}}

\newcommand{\ggabs}{\gamma\gamma}

\begin{document}
\title{On the multiwavelength emission from Gamma Ray Burst afterglows}
\author{M. Petropoulou \and A. Mastichiadis
}
\institute{Department of Physics, University of Athens, Panepistimiopolis, GR 15783 Zografou, Greece}
\offprints{A. Mastichiadis
}
\date{Received ... / Accepted ...}
\abstract{}{Drawing an analogy with Active Galactic Nuclei,
we investigate the one-zone SSC model of Gamma Ray Bursts afterglows
in the presence of electron injection and cooling both by synchrotron and SSC losses.}
{We solve the spatially
averaged kinetic equations which describe the simultaneous evolution of particles and photons, obtaining the multi-wavelength spectrum as a
function of time. We back up our numerical calculations with analytical solutions
of the equations using various profiles of the magnetic field evolution
under certain simplifying assumptions.}{We apply the model to the afterglow evolution of GRBs
in a uniform density environment
and examine the impact various parameters have on the multiwavelength spectra.
We find that in cases where the electron injection and/or the ambient density is
high, the losses are dominated by SSC and the solutions depart significantly from
the ones derived in the synchrotron standard cases.}{}

%
%
\maketitle

\section{Introduction}

Gamma Ray Burst (GRB) afterglows are thought to be produced in the
Relativistic Blast Waves (RBW) associated with the initial GRB
explosion. According to the standard model (for a review, see
\citet{piran05}), the RBW expands in the circumburst material
and, after sweeping some critical amount of mass, it starts
decelerating. At the same time electrons are postulated to energize
at the shock front  and radiate through synchrotron
\citep{SariPiran98,dermchiang98} or synchrotron and SSC
\citep{chiangderm99,fan2008,waxman97a,kumar_panaitescu2000}, thus
producing the afterglow emission.

The above picture carries certain analogies to the radiation models
put forward during the last decade to explain the multiwavelength
(MW) observations of blazar emission, and, especially, to the
'one-zone' SSC models (for recent reviews see \cite{bottcher07} and
\cite{mastichiadis09}). These models are based on solving an
equation for the  electron distribution function including
synchrotron and SSC losses and, at the same time, calculate the
radiated photon spectrum. This approach allows for time dependency
to be taken explicitly into account (and thus it can address, for
example, blazar flaring), it can treat the non-linear cooling
associated with SSC and, moreover, it is self-consistent.

Motivated by these developments, we have applied the above technique
to the GRB afterglows. Aim of the present paper is not to fit
spectra or lightcurves, but to focus on the impact the various
parameters, customarily used by researchers in the GRB  field, have
on the MW spectra. We present also analytical solutions of the
electron equation, under certain simplifying assumptions, for a
power-law electron injection suffering synchrotron losses. This,
amongst others, allowed us to test the numerical code. For
definitiveness we restrict our analysis to the uniform density case.

The present paper brings certain improvements over past efforts in the field
\citep{chiangderm99}. First, in the analytical solutions we consider
both standard and non-standard magnetic field evolution. We also use
the full emissivity for synchrotron radiation  instead
of the $\delta-$function approximation. As far as the numerical
part is concerned, we show that the inclusion of SSC both as
electron energy loss mechanism and radiation process can bring,
under certain circumstances, significant departures from the
standard solutions which include only synchrotron radiation as an
energy loss mechanism. Finally, we have included for the first time,
to the best of our knowledge, $\ggabs$ absorption
both as a $\gamma-$ray attenuation and particle reinjection mechanism.

The  paper is structured as follows. In \S2 we review the
basic hydrodynamics and radiative concepts of the GRB afterglow
adopted by the extensive literature on the subject. We also make an
analogy to the blazar case. In \S3 we present the analytical results
for various profiles of the magnetic field. In \S4 we present the
numerical code and the tests we have performed to check its validity. In
\S5 we show various numerical results and we conclude in \S6 with a
summary and a discussion.

\section{Physics of the Relativistic Blast Wave}

\subsection{Hydrodynamics}

 We assume a shell of material with initial mass $M_{0}$ moving with initial bulk Lorentz factor ${\Gamma}_0$.
This will sweep ambient matter and will start decelerating with a
rate determined by energy and momentum conservation. The resulting blast wave
is modeled as having a
 cross sectional area, $A(r)$, that depends on the distance (measured
 in the frame of the explosion). In our work we consider a spherical blast
 wave with $A(r)=4{\pi}r^2$ sweeping a constant density ambient matter. We
 also assume that the bulk kinetic energy which is converted to
 internal energy is not radiated away, contributing to the inertia of
 the blast wave (non-radiative limit). In this case the deceleration
 of the blast wave is determined by a pair of ordinary differential
 equations \citep{blandMckee76}:
\begin{eqnarray}
\label{gamma1}
\frac{d{\Gamma}}{dr} & = &-\frac{A(r){\rho}(r)({\Gamma}^2(r)-1)}{M(r)}\\
\frac{dM}{dr}& = &{\rho}(r)A(r){\Gamma}(r)\quad ,
\end{eqnarray}
where $\Gamma=\Gamma(r)$ is the bulk Lorentz factor of the material,
$M=M(r)$ is the total mass including internal kinetic energy and
$\rho=\rho(r)$ is the mass density of the ambient matter.
Note that Eqn.~(2) implies a non-radiative blastwave evolution. We will 
adopt this assumption throughout the paper.

 It can be shown \citep{blandMckee76} that the above system has an
 analytic solution given by:
\begin{eqnarray}
{\Gamma}(r)=\frac{{\lambda}(x^3-1)+{\Gamma}_0}{(1+2{\Gamma}_0{\lambda}(x^3-1)+{\lambda}^2(x^3-1)^2)^{1/2}}
\quad,
\end{eqnarray}
where ${\lambda}=\frac{4{\pi}{\rho}_0{r_0}^3}{3M_0}$,\quad
$x=\frac{r}{r_0}$ and  $M_0=\frac{E_0}{{\Gamma}_0c^2}$. Here $r_0$
is the initial radius of the blast wave and $\rho_{0}$ is the mass
density of the uniform ambient matter. The evolution of the bulk
Lorentz factor can be separated into three regimes. The first regime
corresponds to the initial period of free expansion of the blast
wave, during which $\Gamma(r)\approx\Gamma_{0}$. The decelerating
phase follows, where the bulk Lorentz factor can be modeled as
\begin{eqnarray}\label{decel}
{\Gamma}(r)=\frac{{\Gamma}_0}{2}\left(\frac{r}{R_d}
\right)^{-3/2}=\sqrt{\frac{{\Gamma}_0}{4{\lambda}}}\left(\frac{r_0}{r}\right)^{3/2}=\tilde{{\Gamma}}r^{-3/2}
\end{eqnarray}
where \eqb\label{gammaconst} \tilde
\Gamma=\frac{\Gamma_0}{2}R_{d}^{3/2} \eqe
 and
 \eqb\label{Rd}
R_d=\left(\frac{3E_0}{4{\pi}nm_pc^2{\Gamma}^2_0}\right)^{1/3} \eqe
is the deceleration radius of the blast wave \citep{ReesM92}. This
power-law $r^{-3/2}$ dependence of the bulk Lorentz factor is often
quoted for a non-radiative blast wave decelerating in a uniform
medium. During the third and final regime the blast wave is practically
non-relativistic.\\
Energy conservation gives the rate of accreted kinetic energy in the
lab frame: \eqb \label{rateen}
\frac{dE}{dt}=c^3A(r)\rho(r)\beta(\Gamma^2(r)-\Gamma(r)) \eqe
This expression follows from the equation of motion (eqn.
 (\ref{gamma1}))
and is applied regardless of whether the blast wave is in the
radiative or adiabatic regime. As the rate of energy accreted is a
Lorentz invariant, expression (\ref{rateen}) holds also in the
comoving frame of the blast wave.

\subsection{Radiation}

The RBW does not only sweep matter and decelerates, as was discussed
above, but it is assumed to be able to energize particles as well.
While there are no detailed models as yet to explain the way
particles achieve high energies, it is assumed that an ad-hoc
fraction $\ee$ of the accreted kinetic energy is injected into
non-thermal electrons with a power-law form. A second assumption
concerns the lower and upper cutoffs of the electron distribution
$\gammamin$ and $\gammamax$. Since the normalization $\tilde q_0$ of
the electron injection is set by the relation
 \eqb \label{normal}
\int_{\gammamin}^{\gammamax}\tilde
q_0\gamma^{-p}(\gamma-1)m_ec^2d\gamma=\ee{{dE}\over{dt}}, \eqe where
$\gamma$ is the Lorentz factor of the electrons and $p$ is the
electron injection spectral index, it is evident that for $p>2$ (as
is typically assumed), only the lower limit $\gammamin$ will play a
role in determining the integral in Eq.(\ref{normal}) and thus
$\tilde q_0$. Therefore the choice of $\gammamin$ has important
consequences on the results. This can be at least of the order of
$\Gamma$ in the comoving frame \citep{mastichiadiskazanas09} but in
case where the electrons are in equipartition with protons it can be
a factor of $(m_p/m_e)$ higher, i.e. $\gammamin=(m_p/m_e)\Gamma$
\citep{katzpiran97, panaitescumeszaros98}.

Since the energetic electrons will emit synchrotron radiation, a
prescription for the magnetic field is also required. This again is
a source of major uncertainty. The usual assumption is that the
magnetic field is in some type of equipartition with the particles;
this implies that the magnetic energy density takes a fraction $\eB$
of the mass accumulated in the RBW, so the magnetic field is given
by the relation \eqb \label{Beq}
 B=({32\pi nm_{\rm p} \eB c^2})^{1/2}\Gamma.
\eqe However one can consider different types of behavior, like
$B\simeq r^{-1}$
\citep{vlahakis2003}.

Depending on the value of the magnetic field, the electrons can cool
(i.e. radiate all their energy) in a dynamical timescale or remain
uncooled, (i.e. keep their energy). These two cases have been called
'fast' and 'slow' cooling \citep{SariPiran98}. From the standard
solutions of the electron kinetic equations
\citep{kardashev62} it is known that cooled electrons have steeper
distribution functions than uncooled ones, i.e. cooled electrons
have an energy dependance that is proportional to $\gamma^{-p-1}$,
while uncooled ones have a $\gamma^{-p}$ dependance, i.e. they still
retain the spectrum at injection.

High energy electrons can also lose energy by inverse Compton scattering on ambient
photons. These photons can illuminate the source of electrons externally
or they can
be the synchrotron photons mentioned earlier. In this latter case the process
is called synchro-self Compton (SSC). The inclusion
of the inverse Compton process has two important
results: (a) It will produce a high
energy spectral component in the photon spectrum and (b) depending
on the respective magnetic and photon field energy densities, it
can alter the electron distribution function, thus affecting  directly the shape of the radiated photon spectrum.

\subsection{Comparisons to blazar models}

The radiation coming from the  GRB shock, as was described above,
has the same underlying physical principles with the so-called
'one-zone' SSC AGN leptonic models set forward to
explain the MW spectrum of blazar emission \citep{inouetakahara96,
mastichiadiskirk97}.
These models, based on earlier ideas set by \cite{maraschietal92} and
\cite{bloommarscher96},
address essentially the same problem of
electron injection, cooling and photon radiation. However
they do not take a 'ready' electron distribution
but, rather, they obtain it from the solution of a kinetic equation
which contains injection, radiative losses, physical
escape from the source and possibly reinjection of particles
as secondaries from photon-photon absorption. The electron
equation is coupled to an equation for photons that has the
usual synchrotron and inverse Compton emissivities written
in a way as to match the radiative electron losses. It can also
have extra terms such as synchrotron self-absorption, photon-photon
pair production, etc.

This kinetic equation approach has the
advantage that it is self-consistent, i.e. the power lost
by the electrons is radiated by the photons. Moreover
the photon energy density can be calculated at each instant and
this feedbacks though the SSC losses on the electron equation, thus this approach can treat
the SSC intrinsic non-linearity.


Therefore it would have been instructive for one to construct a
similar approach to model the radiation of GRB afterglows. However,
one should have in mind, that in GRBs, despite the physical
analogies to AGNs, the situation has some obvious differences which
come mainly from the hydrodynamics of the GRB outflows as this was
outlined in 2.1.

1. AGN modeling involves usually stationary states. This means that
the the radius of the source $R_s$, the Lorentz factor $\Gamma$ and
the magnetic field strength $B$ are all considered constant. This
holds even when short flares are modeled in a time-dependent way
\citep{mastichiadiskirk97, krawczynskietal02, katarzynskietal05}; as
we mentioned earlier, in GRBs all of the above are functions of the
distance r from the origin of the explosion.

2. In GRBs once the profile of the external density is set, then
   the injection of electrons has at least an upper limit as $\epsilon_e$ cannot exceed unity.
There is no such constraint for AGNs as the injected power is essentially a free
parameter.

These differences imply that the numerical codes developed for
blazars cannot be used as they are for the GRB afterglows, but they
have to be modified to take the above into account. We will present
such a code in \S4. However we first show some analytical solutions
that we have derived for the coupled hydrodynamic-radiation problem.

\section{Analytical solutions}
In this section we will first present the kinetic equation for the
electron distribution function, then we will show the solutions for
different magnetic field configurations assuming that synchrotron
losses dominate and we will end by using these solutions to derive
the slope of the synchrotron and SSC lightcurves which corresponds
to each B-field configuration considered.

\subsection{Kinetic equation of electrons}\label{elec}
The equation which governs the electron distribution is:
 \eqb\label{kinetic1}
 \frac{\partial\tilde N(\gamma,t)}{\partial
 t}+\frac{\partial}{\partial \gamma}(\tilde b(\gamma,t)\tilde
 N(\gamma,t))=\tilde Q(\gamma,t) \quad ,
\eqe
 where $\tilde N(\gamma,t)$ is the number of electrons having Lorenz factors between $\gamma, \gamma+d\gamma$ at a time $t$ as measured in the comoving frame.
  The second term of the left hand side takes into
account energy losses of the relativistic electrons due to
synchrotron emission and inverse compton scattering in general. The
term on the right hand side of the equation describes electron
injection. Equation (\ref{kinetic1}) can be also expressed in terms
of the distance $r$ measured in the frame of the explosion, instead
of $t$, through $dt\approx\frac{dr}{c\Gamma(r)}$:

\eqb \label{kinetic2} \frac{\partial N(\gamma,r)}{\partial
r}+\frac{1}{c\Gamma}\frac{\partial}{\partial
\gamma}(b(\gamma,r)N(\gamma,r))=\frac{\tilde Q}{c\Gamma}\equiv
Q(\gamma,r) \eqe
 where $N(\gamma,r)$ is now the number of electrons having Lorenz factors between $\gamma,
 \gamma+d\gamma$ at a radius $r$ as measured in the frame of the explosion.
 In order to obtain some analytical results of the
above equation we consider only synchrotron losses. We assume that a
fraction $\ee$ of the accreted kinetic energy is injected into non
thermal electrons. The injection of the relativistic electrons can
be modeled as a power-law with arbitrary choices of minimum
and maximum Lorentz factors in energy:
 \eqb \label{injection}
Q(\gamma,r)=q_0(r)\gamma^{-p}\Theta(\gamma-\gammamin)\Theta(\gammamax-\gamma)\Theta(r-r_0).
\eqe
 At each radius the normalization of the electron distribution
is given by the prescribed fraction of the power available as bulk
kinetic energy.
Equation (\ref{normal}) leads to
 \eqb
 \label{norma2}
 q_0(r)& = &
\ke \frac{(\Gamma^2-\Gamma)r^2}{c\Gamma}
 \eqe
 The constant $\ke$ which appears in equation
(\ref{norma2}) is given by:
\begin{displaymath}
\ke = \left\{ \begin{array}{ll}
 \ee
\frac{4\pi\rho_0c^3}{m_{\rm e}c^2 \left( \frac{\gammamin^{-p+2}-\gammamax^{-p+2}}{p-2}-\frac{\gammamin^{-p+1}-\gammamax^{-p+1}}{p-1}\right)} & \textrm{if $p\neq 2$}\\
\\
\ee\frac{4\pi\rho_0c^3}{m_{\rm e}c^2 \left(\ln
\left(\frac{\gammamax}{\gammamin}\right)+\left(\frac{1}{\gammamax}-\frac{1}{\gammamin}\right)\right)}
& \textrm{if $p=2$}
\end{array} \right.
\end{displaymath}

\subsubsection{Magnetic Field of the form
$B(r)=B_0 r^{-3/2}$}\label{equipartion}

\begin{figure}
\resizebox{\hsize}{!}{\includegraphics[angle=270]{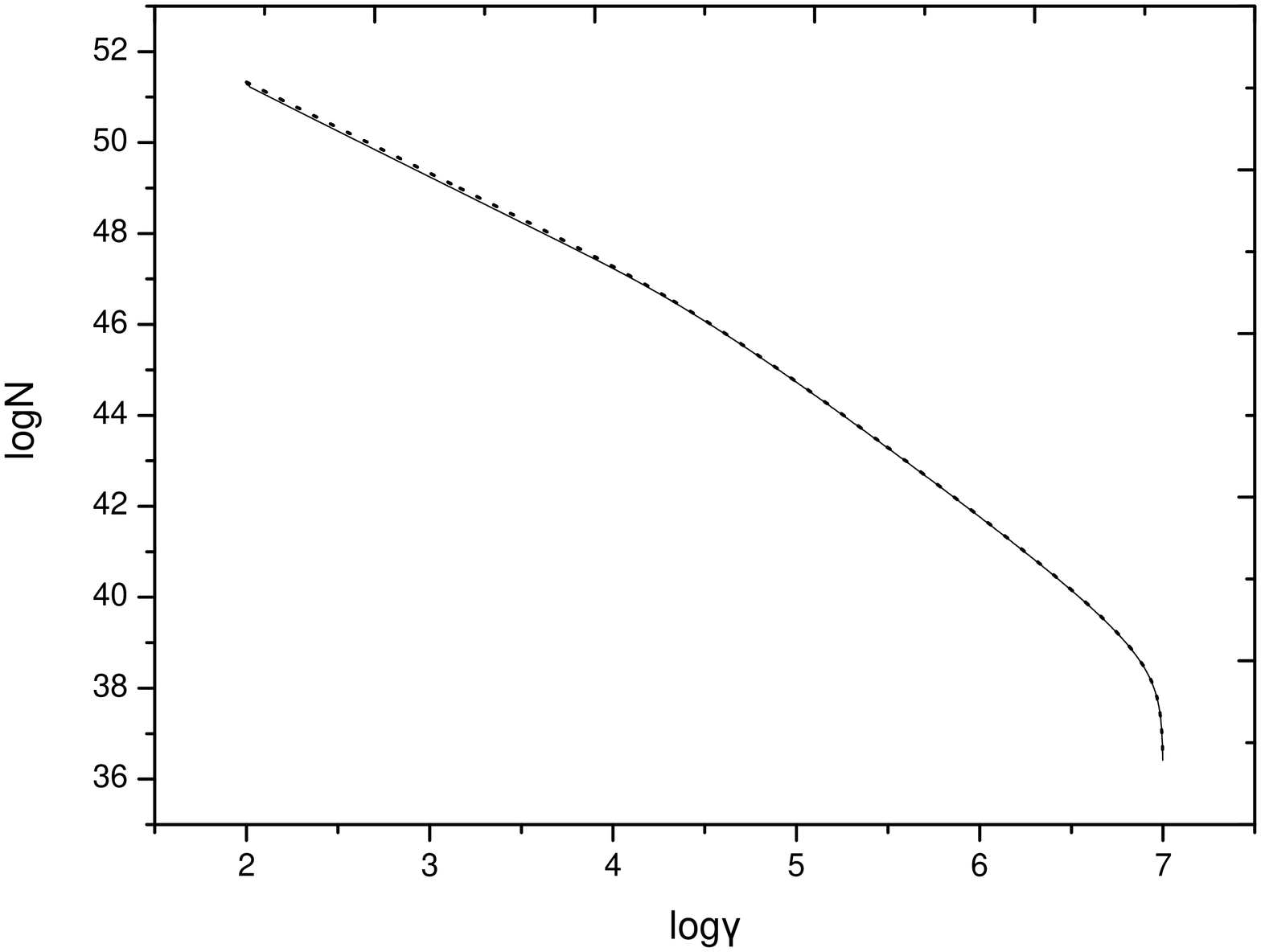}}
 \caption{Comparison between analytical (solid line) and numerical (dotted line) solution
in case of $p=2$ at radius $r =1.5\times 10^{17} \textrm{cm}$ or at
observer time $2.7\times 10^2~s$. There is practically no difference
between the two solutions. The parameters used for this plot are:
 $\epsilon_{\rm e}=0.1, \epsilon_{\rm B}=0.001, n=1
 \textrm{cm}^{-3},\Gamma_{0}=400, E_{0}=10^{53} \textrm{erg}~ \textrm{s}^{-1},\gammamin =10^2,\gammamax =10^{7}$}
 \label{figureN1}
\end{figure}

As was briefly discussed in Section 2, this type of B-field is the
one customarily adopted for the GRB afterglows. Eqn. (\ref{Beq})
implies that a fraction $\eB$ of the accumulated mass on the RBW
goes to amplify the B-field. As $\Gamma\propto r^{-3/2}$ in the
decelerating phase
 (c.f. eq.(\ref{decel}))
the above prescription for the B-field is derived at once. Thus
Eqn.(\ref{Beq}) can be further written
 \eqb
 B=B_{0}r^{-3/2}
 \eqe
 with
 \eqb
 B_{0}=({32\pi nm_{\rm p} \eB c^2})^{1/2} \tilde \Gamma
 \eqe
where the constant $\tilde \Gamma$ is defined in equation
 (\ref{gammaconst}).
Then the term of synchrotron losses becomes
 \eqb
 b(\gamma,r) & = & -\alpha_0\frac{\gamma^2}{r^3}
 \eqe
where $\alpha_0=\frac{\sigma_T B_{0}^{2}}{6 \pi cm_{\rm e}}$. 
The solution of equation (\ref{kinetic2}) has a simple expression in
case of $p=2$
\eqb
 N(\gamma,r)=\frac{2 \ke \tilde\Gamma}{3c\gamma^2}K(\gamma,r)
 \eqe
  where
\begin{displaymath}
K(\gamma,r)=\!\! \left\{ \begin{array}{ll} r^{3/2}-r_{0}^{3/2} &
\!\! \textrm{if $\gamma\leq
\gamma_{\rm b}$}\\ \\
r^{3/2}-\left(\frac{1}{\sqrt{r}}+\frac{c \tilde
\Gamma}{2\alpha_0}\left(\frac{1}{\gamma}-\frac{1}{\gammamax} \right)
\right)^{-3} &\!\!\! \textrm{if $\gamma >\gamma_{\rm b}$}
\end{array}\right.
\end{displaymath}
and
 \eqb \gamma_{\rm
b}=\frac{1}{\frac{1}{\gammamax}+\frac{2\alpha_0}{c\tilde
\Gamma}\left(\frac{1}{\sqrt{r_{0}}}-\frac{1}{\sqrt{r}} \right)} \eqe
We point out that the second branch of the above  solution does not
necessarily describe a cooled electron distribution. Only if the
relation
 \eqb\label{condition}
 \frac{c\tilde \Gamma
\sqrt{r}}{2\alpha_{0}\gamma} \ll 1 \eqe holds, then the electron
distribution can be considered cooled, i.e. $N\propto \gamma^{-3}$. A
more physical approach is to consider an electron with Lorentz
factor $\gamma_{\rm c}$, which cools in a timescale equal to the
dynamical $t_{\rm dyn}\approx \frac{2}{5}\frac{r}{c\Gamma}$
\citep{kumar_panaitescu2000}. This is given by:
 \eqb \gamma_{\rm
c}\approx \frac{6\pi m_{\rm e}c}{\sigma_{\rm T}B^2t_{\rm dyn}} \eqe
 Then, electrons with Lorentz factors
greater than $\gamma_{\rm c}$ cool sufficiently. Condition
(\ref{condition}), which arose mathematically, is equivalent to
$\gamma \gg\gamma_{\rm c}$. This condition and relation
(\ref{condition}) differ only by a factor of $5$. From the above
discussion it is clear that the Lorentz factor $\gamma_{\rm b}$ must
not be confused with $\gamma_{\rm c}$, that determines at which
Lorentz factor the cooling of the distribution becomes dominant. An
example is shown in Figure \ref{figureN1}, where the change of the
slope in the power law begins at $\gamma\approx 5\times10^4$. For
the parameters used  $\gamma_{\rm c}= 5.2\times10^4$ and
$\gamma_{\rm b}=2.8\times10^2$.
 In the case where $p\neq 2$ the integral $I_p$ can only be estimated in
two asymptotic regimes. The complete calculations and approximations
can be found in the Appendix. Here we present the expressions for
the electron distribution in these regimes.
\\ \\
 In the uncooled regime the relation
 $\gamma \ll \gamma_{\rm c}$  holds and the solution takes the form:
\eqb N_{\rm uncooled}& \approx\ &\frac{2 \ke \tilde
\Gamma}{3c}r^{3/2}\gamma^{-p}
 \eqe \\
 As we mentioned above in the cooled regime
the condition $\gamma \gg\gamma_{\rm c}$ holds. Then
 \eqb N_{\rm
cooled}& \approx\ &\frac{ \ke \tilde
\Gamma^{2}}{\alpha_{0}(p-1)}r^{2}\gamma^{-p-1}
 \eqe
 \subsubsection{Magnetic field of the form $B(r)=B_0\frac{r_0}{r}$}
 \label{boverr}
This type of magnetic field might also be related to GRB outflows
(see e.g. \cite{vlahakis2003}). Here we present the analytical
solutions of equation (\ref{kinetic2}) for this type of magnetic
field, while the complete calculations can be found in the Appendix.
Again if $p=2$ the solution has the simple form \eqb
N(\gamma,r)=\frac{2 \ke \tilde \Gamma}{3c\gamma^2}K(\gamma,r)\eqe
 where
\begin{displaymath}
K(\gamma,r)= \left\{ \begin{array}{ll} r^{3/2}-r_{0}^{3/2} &
\textrm{if $\gamma\leq
\gamma_{\rm b}$}\\ \\
r^{3/2}-\left(\sqrt{r}-\frac{c \tilde
\Gamma}{2\alpha_0}\left(\frac{1}{\gamma}-\frac{1}{\gammamax} \right)
\right)^3  & \textrm{if $\gamma >\gamma_{\rm b}$}
\end{array}\right.
\end{displaymath}
and
 \eqb \gamma_{\rm
b}=\frac{1}{\frac{1}{\gammamax}+\frac{2\alpha_0}{c\tilde
\Gamma}\left(\sqrt{r}-\sqrt{r_0} \right)} \eqe The constant
$\alpha_{0}$ is given by
 \eqb \alpha_0=\frac{\sigma_T
B_{0}^{2}r_{0}^{2}}{6 \pi cm_{\rm e}}\eqe
 If $p>2$ then the solution
is found in two regimes , as in
(\ref{equipartion}).\\ \\
The uncooled part of the electron distribution is given by:
 \eqb N_{\rm uncooled} \approx\
\frac{2 \ke \tilde \Gamma}{3c}r^{3/2}\gamma^{-p}
 \eqe
which is exactly the same as the one calculated for the magnetic
field $B\propto r^{-3/2}$. In this regime the cooling timescale of
electrons is much greater than the dynamical timescale. Thus, this
part of the electron distribution will not be affected by a
different type of magnetic field which is the cause of electron
cooling.\\ \\
The situation is different for the part of the distribution where
cooling is dominant:
 \eqb
 N_{\rm cooled}  \approx  \frac{ \ke \tilde
\Gamma^2}{\alpha_0(p-1)}r\gamma^{-p-1}
 \eqe
 \subsubsection{Constant magnetic field $B_{0}$}
 \label{constant}
In order to check our analytical results we have also solved
equation (\ref{kinetic2}) for the case of constant magnetic field.
This calculation has already been done by \citet{dermchiang98}
 and thus we can compare our results with
theirs. The outline of the comparison can be found in the
Appendix (\ref{appendix_const}).
In the uncooled regime the solution is given by:
 \eqb N_{\rm uncooled} \approx\
\frac{2 \ke \tilde \Gamma}{3c}r^{3/2}\gamma^{-p}
 \eqe
The above expression for the uncooled part of the electron
distribution is again the same as for the other types of magnetic
field presented in the previous sections. In the cooled regime

 \eqb
 N_{\rm cooled}  \approx  \frac{ \ke \tilde
\Gamma^2}{\alpha_0(p-1)}r^{-1}\gamma^{-p-1}
 \eqe
 In this case the cooled part of the distribution reduces
as the radius of the blast wave increases. This behavior of the
cooled part of the distribution differs from the one presented in
the previous sections where the total number of electrons within the
shell increased with increasing radius.
 \subsection{Analytic flux time profiles}
The kinetic equation of the electron distribution is being solved in
the comoving frame, as shown in section (\ref{elec}). Synchrotron
and SSC spectra are also first calculated in the comoving frame and
then transformed into the observer frame. For this we use a relation
which connects time in the observer frame and radius $r$ which
appears in all our analytical solutions. Thus,
 \eqb
 t_{\rm obs}& \approx & \int_{r_{0}}^{r}\frac{dr}{2c\Gamma^2}
 \eqe
 where $r$ is the radius of the blast wave measured in the comoving
 frame.
 If the
distance of the source from the observer is $D$ then the respective
synchrotron and SSC fluxes at the observer frame are given by: \eqb
\label{synchroflux}
 F_{\rm s}(\nu_{\rm obs},t_{\rm obs}) =
\frac{J_{\rm s}(\nu_{\rm obs},t_{\rm obs})\Gamma}{4\pi D^2} \eqe
\eqb \label{sscflux} F_{\rm ssc}(\nu_{\rm obs},t_{\rm obs}) =
\frac{J_{\rm ssc}(\nu_{\rm obs},t_{\rm obs})\Gamma}{4\pi D^2}
 \eqe
 where $J_{\rm s},J_{\rm ssc}$ are the synchrotron and SSC power per
 unit frequency emitted in the comoving frame.
It is interesting to examine the dependence of the observed fluxes
on time, in cases where the electron kinetic equation can be
analytically solved. Thus, we consider the cases which we have
treated in (\ref{elec}). Since we are interested only in the time
dependency, in what follows we can work using proportionalities.\\
Equations (\ref{synchroflux}) and (\ref{sscflux}) can be reduced to
the simplified form
\eqb
 F_{\rm s}  \propto  C_{\rm
e}\Gamma^{1+\alpha}B^{\frac{p+1}{2}} \eqe
 \eqb
 F_{\rm ssc}  \propto \frac{C_{\rm
e}^{\phantom{e}2}}{r^2}\Gamma^{1+\alpha}B^{\frac{p+1}{2}}
 \eqe
 where $C_{\rm e}=C_{\rm e}(r)$ is the normalization factor of the electron
 distribution.
 After the deceleration
radius (eq.\ref{Rd}) it is straightforward to show that $ t_{\rm
obs}\propto r^{4}$. Thus the synchrotron and SSC fluxes can be
expressed as power laws of the observed time with exponents
$\alpha_{\rm s},\alpha_{\rm ssc}$ respectively. The exponents are
found in the asymptotic regimes of uncooled and cooled electron
distribution and are presented in Table~\ref{table:1}.
\begin{table}
\caption{Negative of slopes of Time Profiles}
 \label{table:1}
\begin{tabular}{|l| l l l|}
\hline
                                  & $B=const$           & $B\propto r^{-1}$       & $B\propto\ r^{-3/2}$\\
\hline
\phantom{1}& \phantom{1}& \phantom{1}& \phantom{1}  \\
$\alpha_{\rm s}^{\rm uncooled}$   & $\frac{3p-3}{16}$   & $\frac{5p-1}{16}$       & $\frac{3p}{8}$     \\
\phantom{1}& \phantom{1}& \phantom{1}& \phantom{1} \\
$\alpha_{\rm s}^{\rm cooled}$     & $\frac{3p+10}{16}$  & $\frac{5p+4}{16}$   & $\frac{6p+1}{16}$     \\
\phantom{1}& \phantom{1}& \phantom{1}& \phantom{1} \\
$\alpha_{\rm ssc}^{\rm uncooled}$ & $\frac{3p-1}{16}$   &$\frac{5p+1}{16}$   & $\frac{3p+1}{8}$\\
\phantom{1}& \phantom{1}& \phantom{1}& \phantom{1} \\
$\alpha_{\rm ssc}^{\rm cooled}$   & $\frac{3p+22}{16}$  &$\frac{5p+8}{16}$   & $\frac{6p+1}{16}$\\
\phantom{1}& \phantom{1}& \phantom{1}& \phantom{1} \\
 \hline
\end{tabular}
\end{table}
The slopes of the synchrotron flux time profiles in the
non-radiative regime (see Table \ref{table:1}) coincide with the
respective ones presented by \cite{dermchiang98}. Table
\ref{table:1} shows that SSC flux time profiles are steeper than the
respective synchrotron profiles both in the uncooled and cooled
regime, for all the magnetic field configurations discussed in
section \ref{elec}.

\section{Numerical approach}
\label{numerical}

\subsection{The code}

The equation solved in the previous section forms the basis of the
electron kinetic equation; we proceed now to augment this with more
processes and to solve it numerically. Since SSC losses depend on
the synchrotron photons energy density, we have to write an
accompanying equation for photons which is coupled to the electron
equation. Similar type of equations have been solved for the blazar
\citep{mastichiadiskirk97}
and prompt/early afterglow GRB cases
\citep{mastichiadiskazanas09} -- note that in this latter case a
third equation for protons was added.

Assuming, as before, that the electrons are a function of distance
from the center of the explosion and energy, their equation reads

\eqb\label{eq1} {\partial\nelec\over\partial r}
+{\cal{L}}^{\mathrm{syn}}_{e}+{\cal{L}}^{\mathrm{ics}}_{e}
+{\cal{L}}^{\mathrm{ad}}_{e} = {\qcal}^{\mathrm{inj}}_{e} +
{\qcal}_{e}^{\mathrm{\gamma\gamma}}
 \eqe while the corresponding
photon equation is

\eqb\label{eq2} \frac{\partial\nph}{\partial
r}+\frac{c\nph}{\tcross}+\cal{L}^{\mathrm{\gamma\gamma}}_{\gamma}+
{\cal{L}}^{\mathrm{ssa}}_{\gamma}={\qcal}^{\mathrm{syn}}_{\gamma}+{\qcal}^{\mathrm{ics}}_{\gamma}.
\eqe

The operators ${\cal{L}}$  denote losses and escape from the
system while ${\cal{Q}}$ denote injection and source terms. The
unknown functions $\nelec$ are $\nph$ are the differential number
densities of electrons and photons respectively and the physical
processes which are included in the kinetic equations are: (1)
electron synchrotron radiation and synchrotron self absorption
(denoted by the superscripts "syn" and "ssa" respectively); (2)
inverse Compton scattering ("ics"); (3) photon-photon pair
production ("$\gamma\gamma$") and (4) adiabatic losses ("adi").

The numerical code keeps the same philosophy, as far as the physical
processes are concerned, with the one described in
\cite{mastichiadiskirk95} (hereafter MK95). However since various
modifications have been introduced, we summarize briefly the
expressions used:

{\sl {(i) Synchrotron Radiation:}} (a) The electron loss term
${\cal{L}}^{\mathrm{syn}}_{e}$ is given by expression (34) of MK95.
(b) The photon emissivity term ${\qcal}^{\mathrm{syn}}_{\gamma}$ is
given using the full high energy emissivity term \citep[see,
e.g.][]{blum70} instead of the delta-function approximation used in
MK95.

{\sl {(ii) Synchrotron-Self Absorption:}} (a) The photon absorption
term ${\cal{L}}^{\mathrm{ssa}}_{\gamma}$ is used as in MK95
(Eqn.~39). (b) There is no matching term for electron heating due to
this process. However, as synchrotron self absorption is expected to
be minimal for the parameters which are of interest here, the error
introduced by this omission is expected to be negligible.

{\sl {(iii) Inverse Compton scattering:}} (a) The electron loss term
${\cal{L}}^{\mathrm{ics}}_e$ is given by solving Eqn. (5.7) of
\citet{blum70}. (b) The photon emissivity term
${\qcal}^{\mathrm{ics}}_{\gamma}$ uses relation (2.48) of
the same paper.

{\sl {(iv) Photon-photon pair production:}} (a) The electron
injection term ${\qcal}_{e}^{\mathrm{\gamma\gamma}}$ is given by
expression (57) of MK95. (b) The photon absorption term
$\cal{L}^{\mathrm{\gamma\gamma}}_\gamma$ is given by expression (54)
of MK95.

{\sl {(v) Electron injection :}} The quantity
${\qcal}^{\mathrm{inj}}_e$ is the electron injection rate which can
take any functional form of distance $r$ and energy $\gamma$.
Following the usually assumed case, we take it to be of a power-law
form as given by Eqn (\ref{injection}). The power-law index, the
normalization and the upper and lower energy cutoffs can be treated
as free parameters.

{\sl {(vi) Photon escape:}} This is characterized by the
light crossing time of the source $t_{\rm cr}=R_{\rm s}/c$ in the
comoving frame. This is not constant but changes according to the
relation $R_s=r/\Gamma$.

There is still one more free parameter to be addressed for
specifying the code quantities and this is the behavior of the
magnetic field with radius. While it is trivial to adopt any
r-dependence we will use the standard case $B\propto r^{-3/2}$
except in one of the tests that follow.

\subsection{Tests}
\label{tests}

There are various tests that we have performed to test the validity
of the code. As the various rates of the radiative processes have
essentially the same form as the ones used in the past to model AGN
MW emission \citep{mastichiadiskirk97, konopelkoetal03}, they have
been checked many times against the results of   e.g. \cite{coppi92,
katarzynskietal05} and others.

\begin{figure}
\centering
\resizebox{\hsize}{!}{\includegraphics[angle=270]{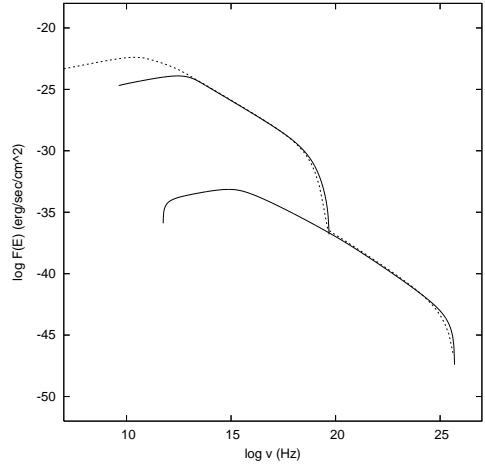}}
\caption{Synchrotron and SSC spectra for a case with $\Gamma_0=100$,
$E_0=10^{53}\textrm{erg}~\textrm{s}^{-1}$, $R_0=10^{14}$cm,
$n=1$~$cm^{-3}$, $p=2$, $\gammamin=10$, $\gammamax=10^4$ and
$B(r)=10^6({R_0\over r})$ Gauss. The spectra are calculated at
$r=2\times 10^{17}$ cm. The source was assumed at a distance of
$D=3~Gpc$. Analytical results are depicted with full line, while the
numerical ones with the dotted one. The difference in the low part
of the spectrum arises from the fact that in the analytical results
we did not solve for the electron distribution below $\gammamin$. In
the numerical case we did not impose such restriction. For
comparison reasons we ignored SSC losses in the numerical code. }
\label{compmaria}
\end{figure}

The new aspect introduced in the code is its dynamical behavior and
in  order to test this we have checked the code extensively against
the analytical results (spectral shape and lightcurve slopes)
derived in the previous section. As an example we show in Figure
\ref{compmaria} such a comparison of  the multiwavelength spectra
derived with the analytical method of the previous section with the
results of the numerical code.

\begin{figure}
\centering
\resizebox{\hsize}{!}{\includegraphics[angle=270]{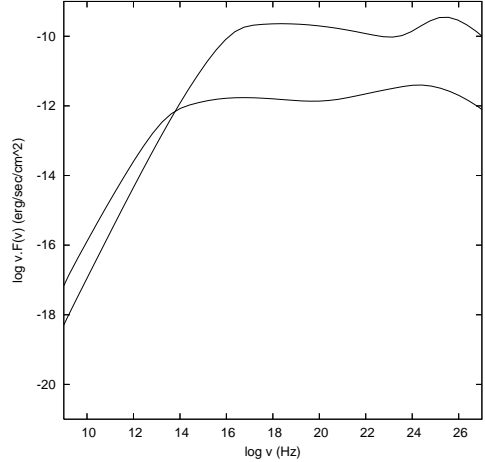}}
\caption{Multiwavelength spectra with parameters similar to the ones
stated in Fig. 4b of Fan et al. (2008). These should be compared
with the two top curves of the aforementioned Figure. }
\label{compfan}
\end{figure}

The other major test was to compare with results already published
in the literature, such as the ones given in \citep{fan2008}. Figure
\ref{compfan} reproduces two curves of Fig. 4b of the aforementioned
paper with very good agreement.

\begin{figure}
\centering
\resizebox{\hsize}{!}{\includegraphics[angle=270]{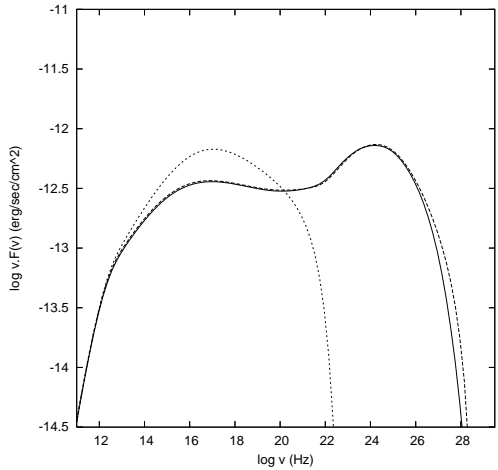}}
\caption{Photon spectrum  obtained at radius $R=3.4\times10^{17}$ cm
or equivalently  at observer time $t_{\rm obs}=7\times 10^3 ~\rm s$
for $\Gamma_0=400$, $E_0=10^{53}\textrm{erg}~\textrm{s}^{-1}$,
$n=1~cm^{-3}$, $\ee=.1$, $\eB=.001$. The electrons were assumed to
have a power-law distribution with slope $p=2.3$ while their maximum
cutoff was $\gammamax=4\times10^7$. $\gammamin$ was given by
expression (\ref{gmin}). The full line corresponds to the spectrum
when all processes are included, the short-dashed one when $\ggabs$
is omitted and the dotted line one when only synchrotron is
included.}

\label{fig4}
\end{figure}

\begin{figure}
\centering
\resizebox{\hsize}{!}{\includegraphics[angle=270]{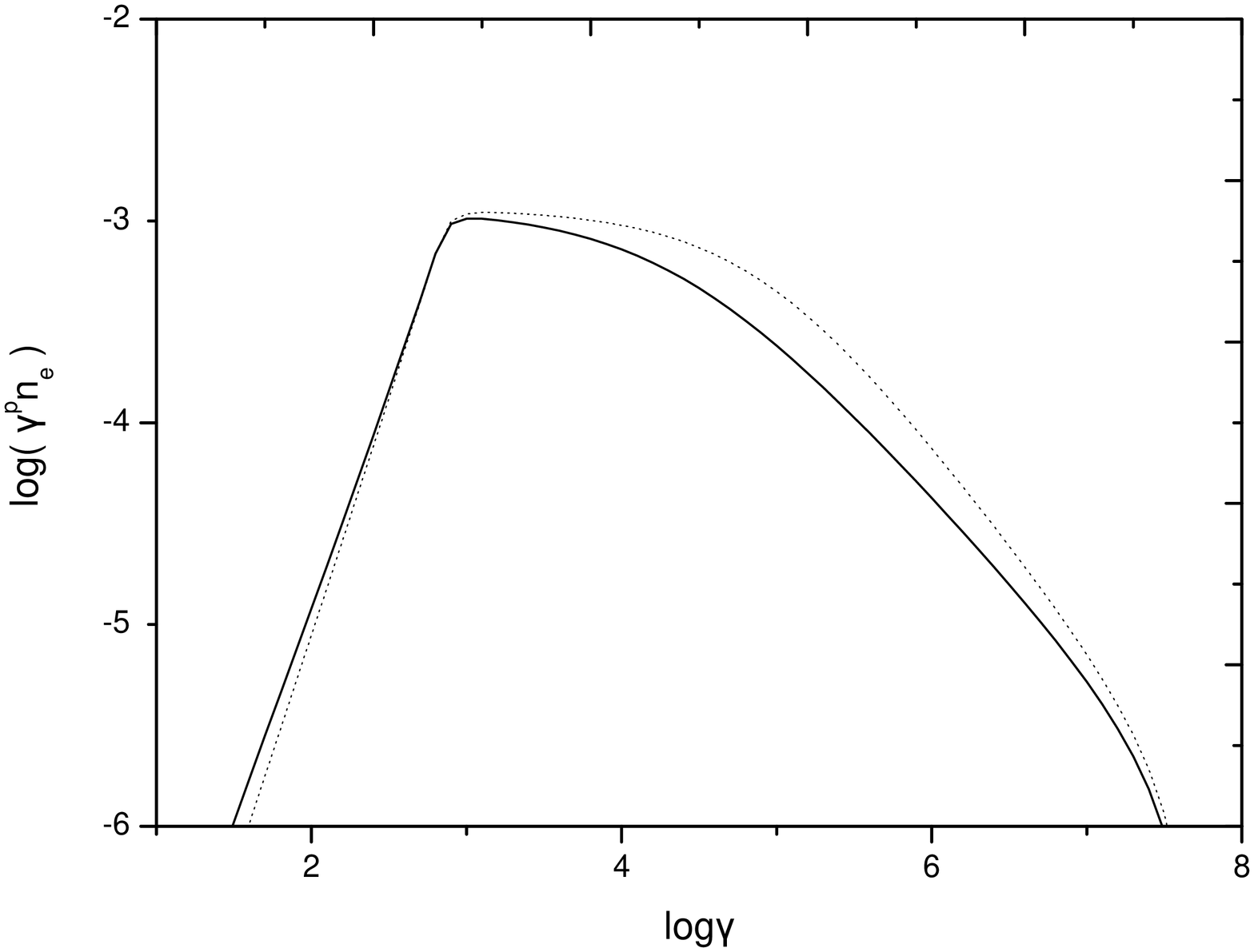}}
\caption{Electron distribution function for $p=2.3$, normalized by
the factor $\sigma_{\rm T}m_{\rm e}c^2R_{\rm s}$ and multiplied by
$\gamma^{p}$, that corresponds to the photon spectrum of Figure
\ref{fig4}. The full line corresponds to the solution when all
processes are included, while the dotted line one when only
synchrotron losses are included.} \label{gN}
\end{figure}

\section{Results}

In what follows we apply the numerical code to examine the dynamical
evolution of the MW GRB afterglow spectra. For this we use the
standard afterglow adopted parameters, i.e. the magnetic field is
given by rel. (\ref{Beq}) with $\eB$ a free parameter, while the
injected electron luminosity is controlled by $\ee$ which is also
treated as a free parameter. Furthermore we adopt a value for
$\gammamin$ that is not constant but is of the form
\citep{SariPiran98} \eqb \label{gmin}
\gammamin={{p-2}\over{p-1}}\ee{{m_p}\over{m_e}}\Gamma \eqe
Our aim
is to see whether (i) SSC losses can modify the electron
distribution function, and therefore, the photon spectrum and (ii)
photon-photon absorption, a process that has been neglected thus
far, can be of some importance, not only taken as a $\gamma-$ray
absorption mechanism but also as an electron (and positron)
reinjection one.

As a first case we show an example for typical values assumed
usually for GRB afterglows. Figure \ref{fig4} shows the photon
spectrum obtained at radius $R=3.4\times10^{17}$ cm for
$\Gamma_0=400$, $n=1~cm^{-3}$, $\ee=.1$, $\eB=.001$. The electrons
were assumed to have a power-law distribution with slope $p=2.3$
while their maximum cutoff was taken to be constant and equal to
$\gammamax=4\times10^7$. Here and in the next Figures the GRB was
set at z=1. Note also that we assume that the evolution
of the Lorentz factor $\Gamma$ follows the adiabatic prescription implied by
Eqn.~(2) and that
we have not taken into account any attenuation for
TeV $\gamma-$rays due to absorption on the IR background. The full
line curve depicts the photon spectrum when all processes are
included, the dashed line one when photon-photon absorption is left
out and the dotted line one when inverse Compton scattering is also
omitted both as emission in the photon equation and as loss
mechanism in the electron equation, i.e. this case can be considered
as pure synchrotron. Although this latter case is clearly an
oversimplification, we have included it for comparison. One can see
that $\ggabs$ absorption influences only the highest part of the
spectrum by making it steeper. Pair reinjection does not alter
significantly the lower spectral parts because only a very small
fraction of the energy has been absorbed and is thus available for
redistribution. On the other hand, SSC losses have an impact on the
spectrum. This effect can be seen better in Figure \ref{gN} which
shows the electron distribution function with (full lines) and
without (dashed lines) SSC losses at the aforementioned radius.
As in standard theory, the braking
energy $\gamma_{\rm c}$ divides the cooled ($\gamma > \gamma_{\rm
c}$) from the uncooled ($\gamma < \gamma_{\rm c}$) part of the
electrons. A first comment one could make is that the synchrotron
break does not appear as a sharp turnover but as a gradual one which
affects the power law index of electrons at least one order of
magnitude around $\gamma_{\rm c}$. A second comment is that
inclusion of SSC losses changes the 'pure' synchrotron picture. The
electron distribution function becomes flatter and this is a result
of the SSC losses. The specific shape can be explained because the
SSC losses at each electron energy consist of both losses in the
Thomson and the Klein Nishina regime. As the electron energy
increases, the fraction of Klein Nishina to Thomson losses also
increases, with the result the total SSC losses to be reduced.

\subsection{Dynamical Evolution}

\begin{figure}
\centering
\resizebox{\hsize}{!}{\includegraphics[angle=270]{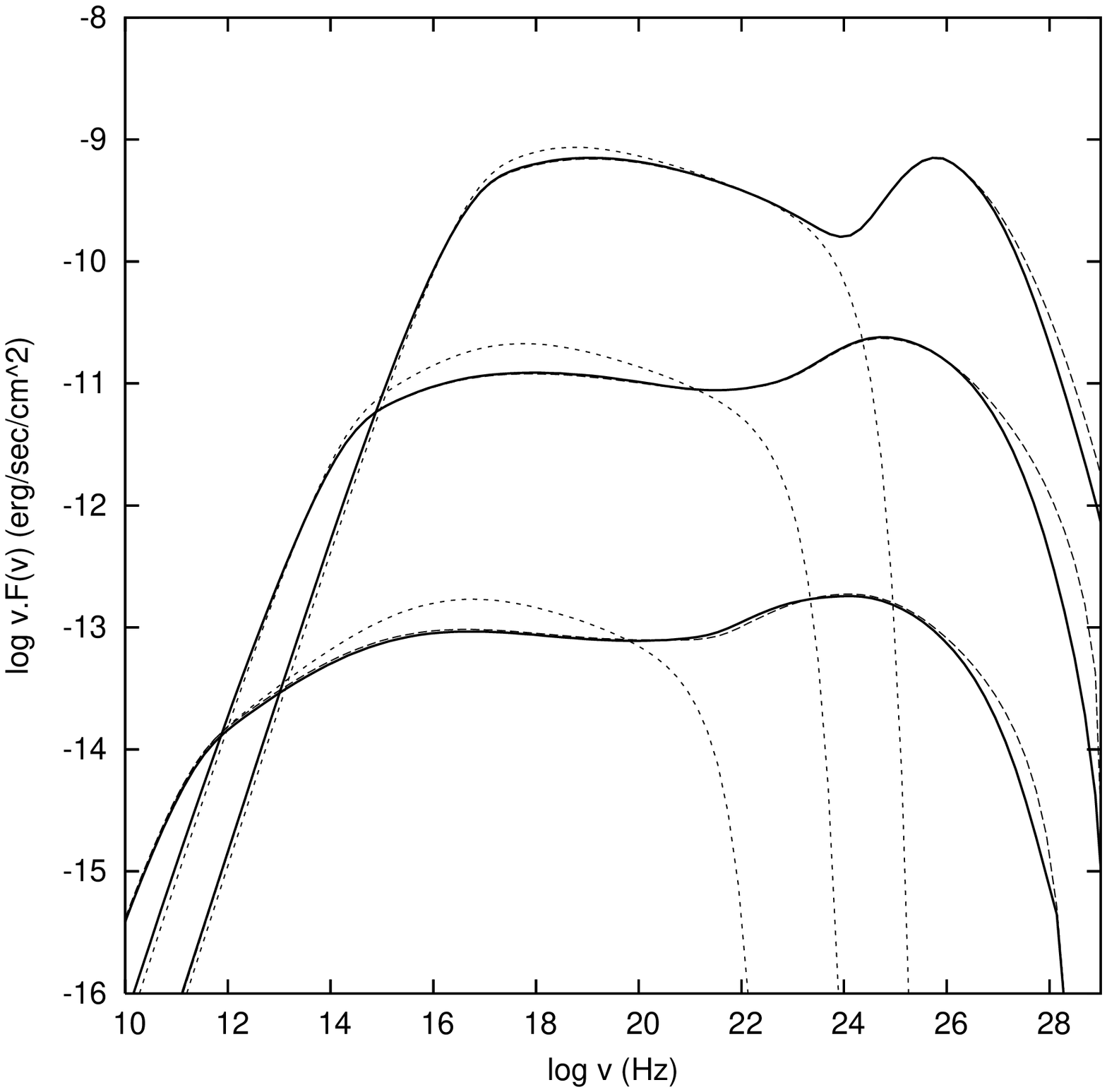}}
\caption{Multiwavelength GRB afterglow spectra for
$\Gamma_0=400$, $E_0=10^{53}\textrm{erg}~\textrm{s}^{-1}$,
$n=1~cm^{-3}$, $\ee=.1$, $\eB=.001$, at $R=\Rdec$, $R=3.2\Rdec$ and
$R=10\Rdec$ (top to bottom) or at equivalent observer times
$t_{\rm obs}=7.2~s$, $2.7\times 10^2~s$, $2.4\times 10^4~s$
respectively. The electrons were assumed to have a power-law
distribution with slope $p=2.3$ and $\gammamax=4\times10^7$. Full
line curves depict the spectrum when all processes are included,
dashed line ones when photon-photon absorption is left out while
dotted line ones are pure synchrotron cases. } \label{MWVR}
\end{figure}

Figure \ref{MWVR} shows snapshots of photon spectra obtained for the
same parameters as above at three different radii: $R=\Rdec$,
$R=3.2\Rdec$ and $R=10\Rdec$. As before, full line curves depict the
spectrum when all processes are included, dashed line ones when
photon-photon absorption is left out and dotted line ones when
inverse Compton scattering is also omitted. Obviously $\ggabs$
absorption affects only the highest energies and does not play any
significant role throughout the evolution.

On the other hand, SSC losses seem to play a role that becomes
slightly more important as the radius increases. This can be
understood from the fact that as the magnetic field drops, two
contradicting results occur: One is that synchrotron cooling becomes
less efficient (it moves from 'fast' to 'slow') and the number of
available soft photons for upscattering is reduced. However, these
same photons become softer with radius (since both $B$ and
$\gammamin$ are reduced outwards) and therefore one expects more
collisions in the Thomson regime where electron losses become more
efficient. This effect can be seen by comparing the shapes of the MW
spectra at the three radii: the shape of the SSC component starts
resembling the synchrotron one as the distance increases and the SSC
losses are dominated by collisions in the Thomson regime.

\subsection{Role of $\ee$}

\begin{figure}
\resizebox{\hsize}{!}{\includegraphics[angle=270]{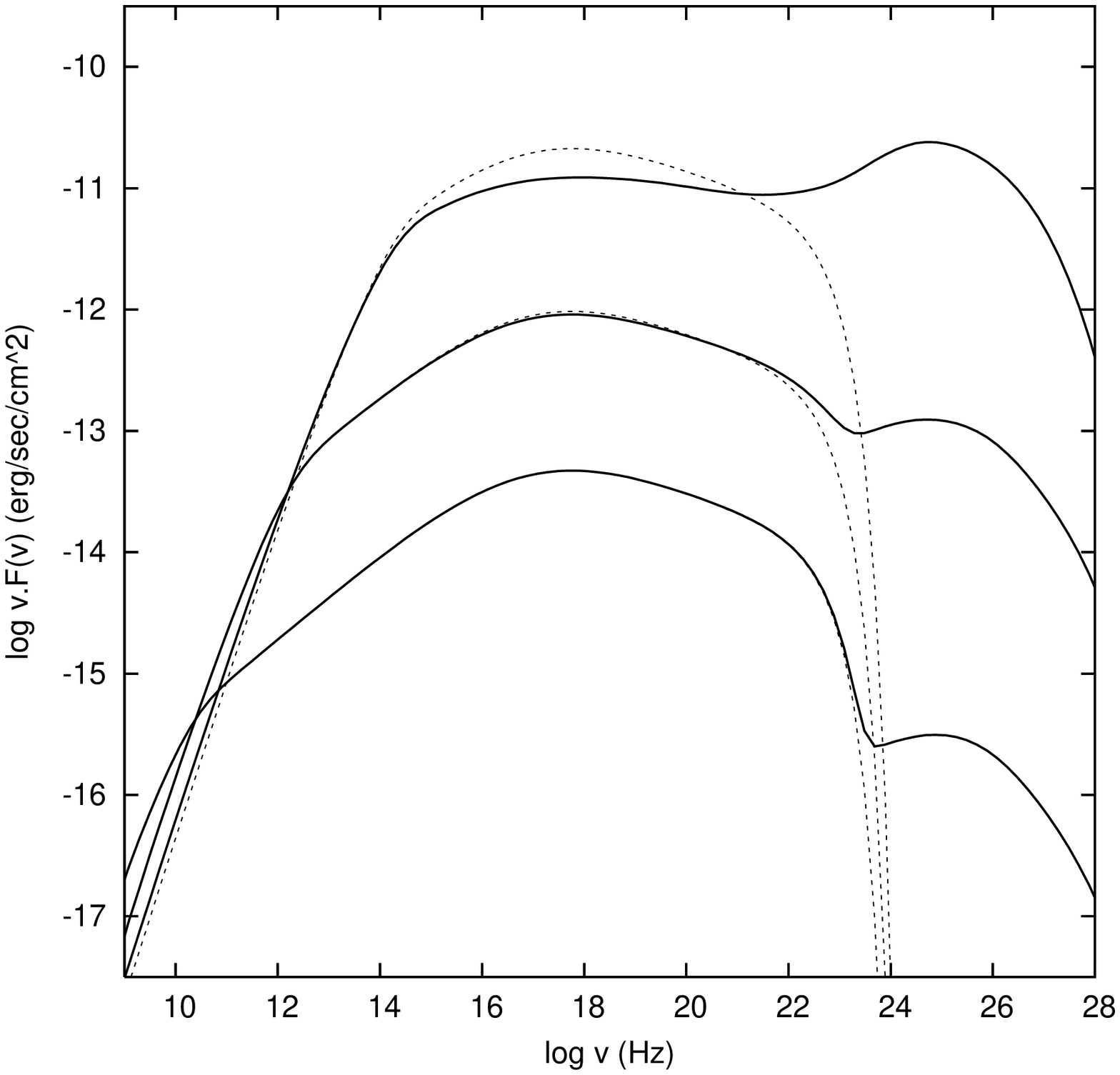}}
\caption{ Multiwavelength GRB afterglow spectra for
$\Gamma_0=400$, $E_0=10^{53}\textrm{erg}~\textrm{s}^{-1}$,
$n=1~cm^{-3}$, $\eB=.001$. The electrons were assumed to have a
power-law distribution with slope $p=2.3$ and
$\gammamax=4\times10^7$. All photon spectra are calculated at the
radius $R=3.2\Rdec$ or at observer time $t_{\rm obs}=2.7\times
10^2~s$ with $\ee=0.1,~0.01$ and $0.001$ (top to bottom). Full line
curves depict the spectra when all processes are included and dotted
line ones depict the corresponding spectra when only synchrotron
radiation is taken into account.} \label{MWVe}
\end{figure}

The effect that $\ee$ has on the spectrum is more straightforward.
As $\ee$ decreases (for fixed $\eB$) the electron spectra are
increasingly dominated by synchrotron losses and the effect of SSC
losses becomes marginal. This is shown in Figure \ref{MWVe} which
depicts the radiated photon spectra for three values of
$\ee=0.1,~0.01$ and $0.001$ (top to bottom). Here $\eB=0.001$, while
the rest of the parameters are as in the previous case. In order to
avoid confusion we have calculated all spectra at radius
$R=3.2\Rdec$. Note that, as $\ee$ decreases, the SSC component drops
as the quadratic of the synchrotron component, a fact that is well
known in the SSC AGN models.

\subsection{Role of $\eB$}

\begin{figure}
\resizebox{\hsize}{!}{\includegraphics[angle=270]{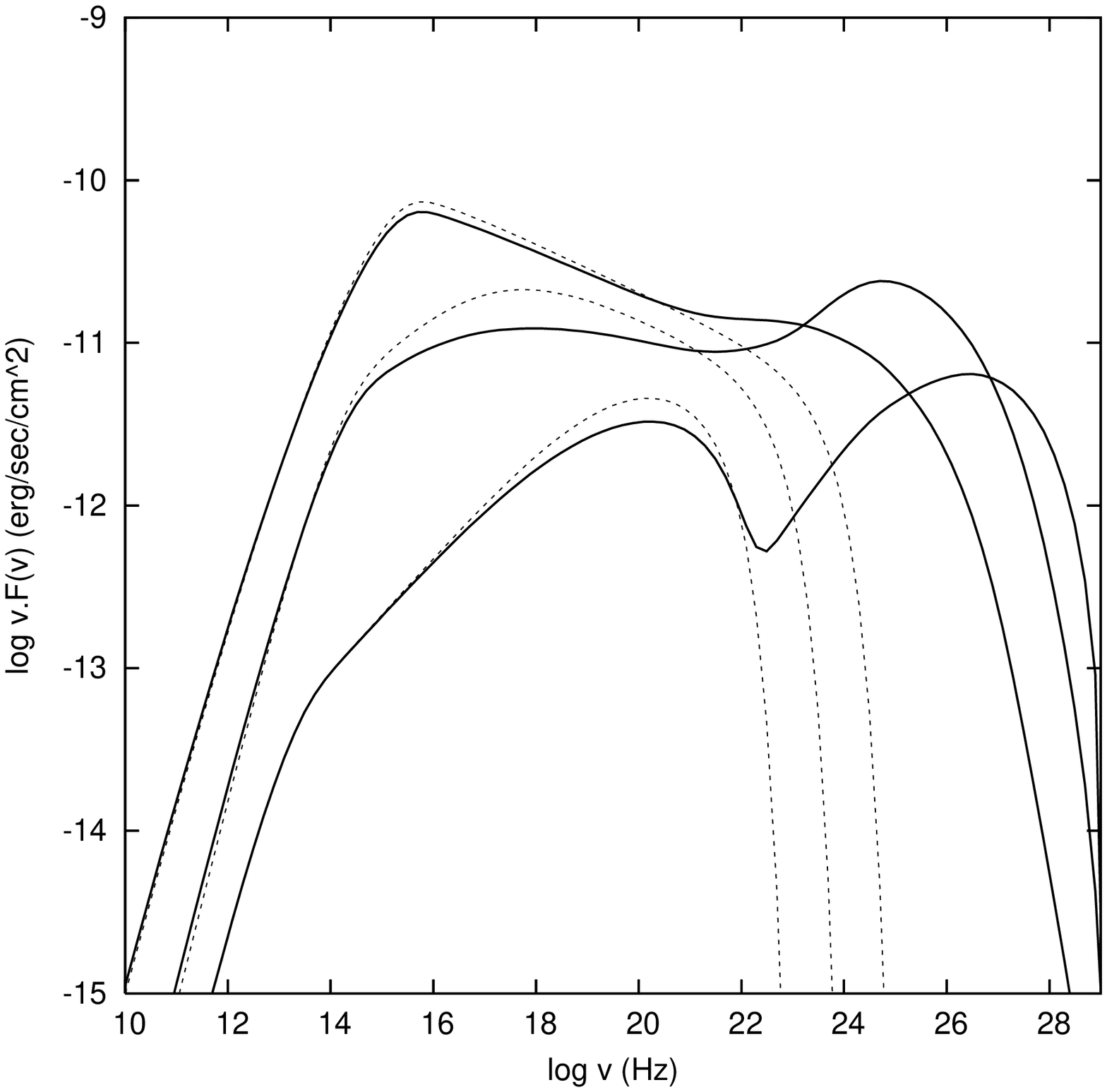}}
\caption{ Multiwavelength GRB afterglow spectra for
$\Gamma_0=400$, $E_0=10^{53}\textrm{erg}~\textrm{s}^{-1}$,
$n=1~cm^{-3}$, $\ee=.1$. The electrons were assumed to have a
power-law distribution with slope $p=2.3$ and
$\gammamax=4\times10^7$. All photon spectra are calculated at the
radius $R=3.2\Rdec$ or at observer time $t_{\rm
obs}=2.7\times10^2~s$ with $\eB=0.1,~10^{-3}$ and $10^{-5}$ (top to
bottom). Full line curves depict the spectra when all processes are
included and dotted line the corresponding spectra when only
synchrotron radiation is taken into account.} \label{figeB}
\end{figure}

Figure \ref{figeB} shows the effects that $\eB$ has on the MW
spectra. Here the run has the same parameters as before, however the
spectra are calculated at the same radius $R=3.2\Rdec$ with
$\eB=0.1,~10^{-3}$ and $10^{-5}$. SSC losses change the spectra only
for intermediate values of $\eB$. High values of $\eB$ lead to fast
cooling and a predominance of synchrotron radiation. On the other
hand, low values of $\eB$ lead to inefficient cooling and only the
high synchrotron frequencies are affected by the SSC cooling which
occurs in the deep KN regime.

\subsection{Role of external density n}

\begin{figure}
\resizebox{\hsize}{!}{\includegraphics[angle=270]{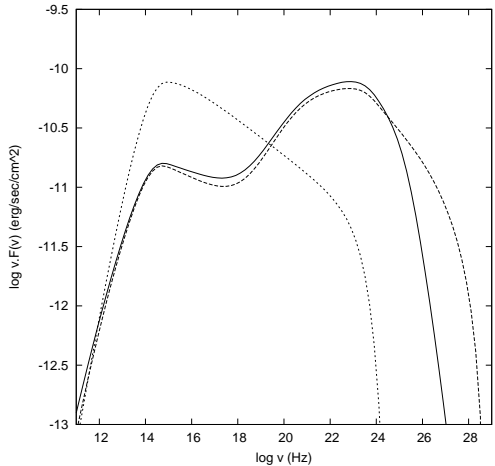}}
\caption{ Multiwavelength GRB afterglow spectra for
$\Gamma_0=400$, $E_0=10^{53}\textrm{erg}~\textrm{s}^{-1}$,
$n=1000~cm^{-3}$, $\ee=.1$ and $\eB=0.001$.  The electrons were
assumed to have a power-law distribution with slope $p=2.3$ and
$\gammamax=4\times10^7$. All photon spectra are calculated at the
radius $R=2.7\times~10^{16}$ cm or equivalently at observer time
$t_{\rm obs}=2.8\times10^2~s$. The full line curve depict the
spectrum when all processes are included, the dashed line one when
photon-photon absorption is left out while the dotted line one is a
pure synchrotron case. Notice that the inclusion of the SSC losses
redistributes the radiated power to high energies  while
$\ggabs$ has an effect on the entire spectrum through
pair injection.} \label{figden}
\end{figure}

Figure \ref{figden} shows the way the MW spectra change in the case
when the density is increased to 1000 part/cm$^{3}$. This figure has
to be directly compared to Fig.~\ref{fig4}. Inclusion of the SSC
losses makes the spectrum to depart significantly from the pure
synchrotron case. Therefore this is a clear Compton dominated case
with the SSC component exceeding the synchrotron one by an order of
magnitude. Furthermore $\ggabs$ absorption produces a
contribution that affects the entire spectrum as substantial
pair injection
redistributes the luminosity from the high energy
end to lower.

\section{Summary/Discussion}
\label{discuss}

In the present paper we have applied the 'one-zone' SSC model which
is customarily applied to the multiwavelength blazar emission to the
GRB afterglows. For this we have used a numerical code that treats
electron injection and cooling and we have calculated
self-consistently the electron distribution and radiated photon
spectrum at each radius of the relativistic blast wave. In this
sense the present work should be considered as complimentary to
\cite{chiangderm99} and \cite{fan2008}.

One difference between the modeling of blazar emission (even in
flaring conditions) and GRB afterglows is that in the latter case
there is continuous evolution of the bulk relativistic Lorentz
factor $\Gamma$, the magnetic field strength $B$ and the radius of
the emitting source $R_s=r/\Gamma$. In order to test the numerical
code in this new setting we have solved analytically the kinetic
equation of the evolving electrons for the uniform density case
under certain simplifying assumptions and compared the resulting
electron distribution function and radiated spectra to the ones
given from the code for similar parameters. The analytical
solutions, details of which can be found in the Appendix, have their
own interest despite the introduced simplifications as they can give
simple expressions for the dependence of the lightcurve on the
spectral index in the case where the losses come solely from
synchrotron. We find, for example, that the magnetic field
prescription plays a significant role in determining the slopes of
the flux time profiles (see Table 1). Moreover they treat correctly the
synchrotron cooling break which should not be taken as an abrupt
change of slope but as a very gradual one.

Restricting ourselves to the uniform density case  in the adiabatic
approximation, we have also
performed numerical runs of the standard GRB afterglow
approach as this can be summarized by the introduction of the usual
parameters $\eB$ and $\ee$. Our aim was two-fold: (i) To see whether
GRB afterglows can run into Compton dominant cases similar to the
ones found in modeling AGNs and (ii) to investigate the effect
$\ggabs$ absorption and subsequent pair reinjection might have on
the afterglow multiwavelength spectra. Starting from the latter case
first we found that, at least for the set of parameters used,
$\gamma\gamma$ absorption has an effect, as a photon attenuation
mechanism, only at the very high energy
part of the spectrum.
Therefore it could affect only the potential TeV
GRB observations -- we point out also that TeV emission from GRBs at
high z will be subject not only to internal $\ggabs$ attenuation,
but also on extra absorption by photons of the IR background, an
effect that already has important consequences even for low z TeV
blazars (see, e.g. \cite{konopelkoetal03})
However, as a pair injection mechanism $\ggabs$ absorption has an effect
over the whole spectrum
that becomes increasingly important with increasing values
of the external density (compare  Fig.\ref{fig4} with
Fig.\ref{figden}).

At the same time, we found that SSC has an
impact on the MW spectra as it introduces an extra source of cooling
on the electron distribution.
As one would expect it depends
also on the parameters $\ee$, $\eB$
and, more critically, on
 the value of ambient density $n$. For low values of $\ee$ it
does not play practically any role and the cooling comes solely from
synchrotron radiation (see Fig.\ref{MWVe}).
Thus the GRB afterglow is in the "synchrotron
dominated" regime.
However for high values
of $\ee$ it becomes important and this can be seen from the fact
that in the $\nu F_\nu$ spectra the SSC component carries about the
same luminosity as the synchrotron component even in the
$n=1~cm^{-3}$ case.
The effect becomes quite severe for higher values
of $n$ where the SSC component carries an increasingly higher part
of the luminosity (see Fig.~\ref{figden}). These afterglows are
clearly Compton dominated.

The fact that we have limited our analysis to the adiabatic case
restricts the allowed values of $\ee$. As a limiting case
we have adopted the value $\ee=0.1$ which introduces a maximum
(corresponding to the fast cooling case)
error of a few percent in the Lorentz factor $\Gamma$
as this has to be
corrected for radiative losses \citep{chiangderm99}.
In any case, these corrections are small, they do not affect
the spectral shapes and, therefore, they cannot alter the basic
results of the present paper.

Concluding we can say that the inclusion of the effects of SSC (not
only as an emission, but as a loss process as well) and, to a
smaller degree, of $\ggabs$ absorption can have a significant impact
on the GRB afterglow multiwavelength spectra. As expected, the
overall picture that emanates from these processes is rather
complicated. However, as a rule of thumb, we can say that the
aforementioned processes become increasingly important for high
external densities and high $\ee$ cases. Their inclusion has some
non-trivial consequences in the lightcurves as both the photon
spectra slope and the energy break are modified. On the other hand,
they can be rather safely neglected for high $\eB$ and low $\ee$ or
$n$ cases.

Based on the above results we expect that SSC losses will play an
important role in a wind-type density profile, especially at early
times when densities are high. This will be the subject of another
paper.

\begin{acknowledgements}
We would like to thank Prof. T. Piran and Dr. D. Giannios for many
interesting discussions and comments on the manuscript.
We also thank the referee, Dr. M. B\"ottcher, for his swift
reply and for making points that helped improve the manuscript.
This research was funded in part by a Grant
from the special Funds for Research (ELKE) of the University of
Athens.
\end{acknowledgements}

\appendix
\section{Analytical solutions of the electron kinetic equation}
\subsection{Magnetic field of the form $B=B_0r^{-3/2}$}
The solution of equation (\ref{kinetic2}) is
 \eqb\label{ne1}
N(\gamma,r)=\int_{-\infty}^{r}d\tilde r Q(\tilde \gamma,\tilde
r)\frac{\tilde \gamma^2}{\gamma^2}\eqe
 or changing the integration variable
 \eqb
 N(\gamma,r)=\frac{1}{\alpha_0
\gamma^2}\int_{\gamma}^{\infty}d\tilde \gamma Q(\tilde \gamma,\tilde
r)c\Gamma(\tilde r) \tilde r^3 \eqe
 The electron distribution is in
general given by:
 \eqb N(\gamma,r)=
\frac{\ke \tilde\Gamma^2}{\alpha_0\gamma^2}I_p \eqe where constant
$\alpha_0$ is defined in section \ref{equipartion}.
 \eqb
\label{Ipeq} I_p & = & \int_{\gamma}^{\rm
min(\gamma_{\ast},\gammamax)} \!\!\!\!\!\!\!\!d\tilde \gamma \tilde
\gamma^{-p} \left[\frac{1}{\sqrt{r}}+\frac{c\tilde
\Gamma}{2\alpha_{0}}\left(\frac{1}{\gamma}-\frac{1}{\tilde
\gamma}\right) \right]^{-4}\\  \nonumber \\
\gamma_{\ast} & =
&\frac{1}{\frac{1}{\gamma}+\frac{2\alpha_0}{c\tilde \Gamma}
\left(\frac{1}{\sqrt{r}}-\frac{1}{\sqrt{r_{0}}}\right)} \eqe We
define $A=\frac{c\tilde \Gamma}{2 \alpha_{0}}$. The integral for
$p>2$ (as it is typically assumed) can be estimated in two regimes:\\ \\
\textit{\large Uncooled}\\ \\
In this regime $\gamma<\gamma_{\rm c}$ and
$\gamma_{\ast}<\gammamax$. Taking also into account that
$\gamma_{\ast} > \gamma$ and $p>2$ equation (\ref{Ipeq}) becomes:
 \eqb
 I_{p}^{\rm uncooled}=-\frac{2\alpha_0}{3c\tilde
\Gamma}\tilde \gamma^{-p+2} \left[\frac{1}{\sqrt{r}}+\frac{c\tilde
\Gamma}{2\alpha_0} \left(\frac{1}{\gamma}-\frac{1}{\tilde\gamma}
\right) \right]^{-3} \Bigg \vert_{\gamma}^{\gamma_{\ast}}
 \eqe
 or
 \eqb
 I_{p}^{\rm uncooled}& \approx & \frac{2\alpha_0}{3c\tilde
\Gamma}r^{3/2}\gamma^{-p+2}\\ \nonumber \\ \label{uncooleq} N_{\rm
uncooled}& \approx\ &\frac{2\ke \tilde \Gamma}{3c}r^{3/2}\gamma^{-p}
 \eqe \\
 \textit{\large Cooled}\\ \\
 This is the regime where the bulk of the electron
population has already cooled and corresponds to the conditions
$\gamma >\gamma_{\rm b}$ and $\gamma \gg \gamma_{\rm c}$ or
equivalently $\frac{A \sqrt{r}}{\gamma}<<1$, as mentioned in section
(\ref{equipartion}). The integral $I_{\rm p}$ is then calculated:
 \eqb \label{Imakrinari}
 I_{\rm p} =-\frac{1}{6}\Big [f_{0}(\gamma,r;p)\big[f_{1}(\gamma,r;p)f_{2}(\gamma,r;p)+f_{3}(\gamma,r;p)\big]\Big]
 \eqe
where
 \eqb
 f_{0} & = & \frac{\gamma^5}{(p-1)\left(A+\frac{\gamma}{\sqrt{r}}\right)^5}
\eqe
 \eqb
 f_{1} & = &
-\frac{1}{\gamma^3}\left(-\frac{A+\frac{\gamma}{\sqrt{r}}}{\gamma}
\right)^{p}\left(-\frac{\sqrt{r}}{\gamma+A\sqrt{r}}\right)^{p-1}
\eqe \eqb
 f_{2}& = &
 -(\gamma+A\sqrt{r})(p-1)\cdot \nonumber \\ & & {} \cdot(2A^2r-A\gamma\sqrt{r}(p-8)+ \gamma^2(18-8p+p^2))
+ \nonumber \\ & & {} + \gamma^3(-24+26p-9p^2+p^3) \cdot \nonumber
\\ & & {} \cdot F\left(1,p-1;p;1-\frac{\gamma}{\gamma+A\sqrt{r}}\right)
 \eqe
\eqb \label{f3} f_{3} & = &
6\gammamax^{-p+1}\left(\frac{A}{\gamma}+\frac{1}{\sqrt{r}}\right)\cdot
\nonumber
\\ & & {} \cdot F\left(4,p-1;p;\frac{A\gamma}{\gammamax\left(A+\frac{\gamma}{\sqrt{r}}\right)}\right)
\eqe where $F(a,b;c;z)$ is the Gaussian hypergeometric function. The
hypergeometric function has a power series representation. The three
first terms of the series are: \eqb F(a,b;c;z)& = &
1+\frac{ab}{1!c}z+ \nonumber \\ & & {} + \frac{a(a+1)b(b+1)}{2!
c(c+1)}z^2+\ldots \eqe The condition $\frac{A \sqrt{r}}{\gamma}<<1$
allows us to expand each term of equation(\ref{Imakrinari}) in terms
of $\frac{A \sqrt{r}}{\gamma}$. We work only in zeroth order, in
order to obtain the simplest expression for $I_{\rm p}$ and its
dependance on $r,\gamma$:
 \eqb
\frac{\gamma^5}{(p-1)\left(A+\frac{\gamma}{\sqrt{r}}\right)^5}
&\!\approx &\! \frac{r^{5/2}}{p-1} \\ \nonumber \\
(-1)^{2p}\frac{1}{\gamma^3}\left(\frac{A+\frac{\gamma}{\sqrt{r}}}{\gamma}\right)^{p}\left(\frac{\sqrt{r}}{\gamma+A\sqrt{r}}\right)^{p-1}\!\!\!\!
&  \approx & \!\!\!
 \gamma^{-p-2}r^{-1/2} \\ \nonumber \\ \label{1}
6\gammamax^{-p+1}\left(\frac{A}{\gamma}+\frac{1}{\sqrt{r}}\right) &
\! \approx & \! \frac{6\gammamax^{-p+1}}{\sqrt{r}}
 \eqe \\
The term of equation (\ref{1}) will not be taken into account later
on. The argument of the hypergeometric function in equation
(\ref{f3}) becomes
 \eqb
\frac{A\gamma}{\gammamax\left(A+\frac{\gamma}{\sqrt{r}}\right)}
\approx  \frac{A\sqrt{r}}{\gammamax}\ll 1 \eqe Thus, we can
approximate \eqb F\left(4,p-1;p;
\frac{A\gamma}{\gammamax\left(A+\frac{\gamma}{\sqrt{r}}\right)}\right)
\approx 1 \eqe The hypergeometric function with argument
$1-\frac{\gamma}{\gamma+A\sqrt{r}}\rightarrow 0$ can also be
approximated by unity. With the above approximations, equation
(\ref{Imakrinari}) reduces  to: \eqb I_{\rm
p}^{\rm cooled} \approx \frac{1}{p-1}r^2\gamma^{-p+1} \eqe\\
Thus, the electron distribution in the cooled regime is given by the
simple expression: \eqb N_{\rm cooled} \approx \frac{\ke \tilde
\Gamma^2}{\alpha_{0}(p-1)}r^2\gamma^{-p-1} \eqe

\subsection{Magnetic field of the form $B(r)=B_0\frac{r_0}{r}$}
The  general solution of equation (\ref{kinetic2}) is
 \eqb\label{ne1}
N(\gamma,r)=\int_{-\infty}^{r}d\tilde r Q(\tilde \gamma,\tilde
r)\frac{\tilde \gamma^2}{\gamma^2}\eqe
 or changing the integration variable
 \eqb
 N(\gamma,r)=\frac{1}{\alpha_0
\gamma^2}\int_{\gamma}^{\infty}d\tilde \gamma Q(\tilde \gamma,\tilde
r)c\Gamma(\tilde r) \tilde r^2 \eqe where constant $\alpha_0$ is
defined in section (\ref{boverr}).
 Assuming that the blast wave is
in the decelerating phase (eq.(\ref{decel})) the above equation
becomes: \eqb N(\gamma,r)= \frac{\ke
\tilde\Gamma^2}{\alpha_0\gamma^2}I_p \eqe where
 \eqb\label{integral}
 I_p & =
& \int_{\gamma}^{\mathrm{min}(\gamma_{\ast},\gammamax)} d\tilde
\gamma \tilde \gamma^{-p} \left[\sqrt{r}-\frac{c\tilde
\Gamma}{2\alpha_0}\left(\frac{1}{\gamma}-\frac{1}{\tilde
\gamma}\right) \right]^2\\
\gamma_{\ast} & =
&\frac{1}{\frac{1}{\gamma}-\frac{2\alpha_0}{c\tilde
\Gamma}(\sqrt{r}-\sqrt{r_0})} \eqe
 If $p\neq 2$ the
integral $I_p$ can only be estimated in two regimes:\\
\begin{itemize}
\item \textit{Uncooled}
\end{itemize}

 In this regime, $\gamma<\gamma_{\rm c}$ and $\gamma_{\ast}<\gammamax$. Equation (\ref{integral})
becomes
 \eqb \label{uncool}
 I_{p}^{\rm
uncooled}=-\frac{2\alpha_0}{3c\tilde \Gamma}\tilde \gamma^{-p+2}
\left[\sqrt{r}-\frac{c\tilde \Gamma}{2\alpha_0}
\left(\frac{1}{\gamma}-\frac{1}{\tilde\gamma} \right) \right]^3
\Bigg \vert_{\gamma}^{\gamma_{\ast}}
 \eqe
Taking into account that $\gamma_{\ast}>\gamma$ and $p>2$ (as it is
typically assumed), inspection of (\ref{uncool}) shows that:
 \eqb
 I_{p}^{\rm uncooled}& \approx & \frac{2\alpha_0}{3c\tilde
\Gamma}r^{3/2}\gamma^{-p+2}\\ \nonumber \\ \label{overrunc} N_{\rm
uncooled}& \approx\ &\frac{2\ke \tilde \Gamma}{3c}r^{3/2}\gamma^{-p}
 \eqe \\
\begin{itemize}
\item \textit{Cooled}
\end{itemize}
 This is the regime where the bulk of the electron
population has already cooled and corresponds to the condition
$\gamma>\gamma_{\rm b}$ or equivalently $\gamma_{\ast}>\gammamax$.
When $\gamma<<\gammamax$
 \eqb
 I_p\rightarrow I_{p}^{\rm cooled}=\frac{\gamma^{-p+1}}{p-1}
 \left(\sqrt{r}-\frac{c\tilde \Gamma}{2\alpha_0\gamma}\right)^2
\eqe Moreover the condition $\gamma_{\ast}>\gammamax$ leads to
$\frac{c\tilde \Gamma}{2\alpha_0 \gamma}<\sqrt{r}$. 
However, we can safely neglect the term
$\frac{c\tilde\Gamma}{2\alpha_{0}\gamma}$, only if it is much
smaller compared with the term $\sqrt{r}$. This condition , as
discussed in section (\ref{equipartion}), is the same as $\gamma\gg
\gamma_{\rm c}$. Under this approximation:
 \eqb
 I_{p}^{\rm cooled}& \approx & \frac{\gamma^{-p+1}}{p-1}r\\ \nonumber \\
N_{\rm cooled} & \approx & \frac{\ke \tilde
\Gamma^2}{\alpha_0(p-1)}r\gamma^{-p-1}
 \eqe

\subsection{Constant magnetic field}
\label{appendix_const}
 The solution of equation (\ref{kinetic2}) is:
\eqb\label{ne1} N(\gamma,r)=\int_{-\infty}^{r}d\tilde r Q(\tilde
\gamma,\tilde r)\frac{\tilde \gamma^2}{\gamma^2}\eqe
 or changing the integration variable
 \eqb
 N(\gamma,r)=\frac{1}{\alpha_0
\gamma^2}\int_{\gamma}^{\infty}d\tilde \gamma Q(\tilde \gamma,\tilde
r)c\Gamma(\tilde r) \tilde r^{-1} \eqe where constant $\alpha_0$ is
defined in section (\ref{constant}).
 As described in the previous
sections , the solution in terms of the integral $I_p$ is:
 \eqb \label{Nours}
 N(\gamma,r)=
\frac{\ke \tilde\Gamma^2}{\alpha_0\gamma^2}I_p \eqe where in this
case
 \eqb \label{Iours}
 I_p & = & \int_{\gamma}^{\rm
min(\gamma_{\ast},\gammamax)}\!\!\!\!\!\!d\tilde \gamma \tilde
\gamma^{-p}\left[r^{5/2}-\frac{5c\tilde
\Gamma}{2\alpha_{0}}\left(\frac{1}{\gamma}-\frac{1}{\tilde
\gamma}\right)\right]^{-2/5} \\ \nonumber \\
\gamma_{\ast}& =
&\frac{1}{\frac{1}{\gamma}-\frac{2\alpha_{0}}{5c\tilde
\Gamma}\left(r^{5/2}-r_{0}^{5/2}\right)}
 \eqe
We outline next the points needed that will faciliate
a comparison of our analytical results and those presented in \cite{dermchiang98} (hereafter DC98).
 The expressions given by equations (\ref{Nours}), (\ref{Iours}) coincide with these of equations (33), (34) of DC98.
 This can be seen after taking into account the corresponding symbolism of our present work with the one used in
 DC98. The spatial coordinates $x,x_0$, used in DC98, and
 $r,r_0$, used in our work, are identical.
In the case of a spherical blast wave  \eqb A_0=4\pi r_0^2. \eqe When
the bulk of the electron distribution is considered to be
relativistic then $\beta\approx 1$ and
 \eqb p=\beta\gamma\approx \gamma.\eqe
The constant $\tilde \Gamma$ used troughout our paper is related to
$\Gamma_0,x_0$ of DC98 by
 \eqb \tilde\Gamma
= \Gamma_0 x_{0}^{3/2}.\eqe
Our constant $k_{\rm e}$ is related
to the constant in the right hand side of Eqn. (33) of DC98
by the relation
\eqb k_{\rm e}=\frac{\xi_{\rm e}4\pi \rho_0 c}{m_{\rm
e}f},\eqe where $\xi_{\rm e}\equiv \epsilon_{\rm e}$ and $\rho_0=
m_{\rm p} n_0$. The relationship between the two integrals
$I_p(\tau)$ in DC98 and $I_p(\gamma,r)$ in our paper is given by \eqb
I_p(\tau)=\frac{r_0}{\alpha_0}I_p(\gamma,r).\eqe
Moreover, the constants
$\nu_0$ and $\omega$ which appear in Eqn. (34) of DC98 can be
expressed as \eqb \nu_0 & \equiv &
\alpha_0 \\
 \omega & = & \frac{5 \tilde \Gamma c}{2r_{0}^{5/2}}.\eqe
Also the variable $\tau$ of DC98 can be expressed in terms of the
distance $r$ measured in the frame of the explosion by
the relation
\eqb \tau =
\frac{2}{5c \tilde \Gamma}\left(r^{5/2}-r_0^{5/2}\right).\eqe
 Finally
the exponent $u$ which appears in the integral $I_p(\tau)$ takes the
value $-\frac{2}{5}$.
 The characteristic Lorentz factor $\gamma_{\rm b}$ which divides the analytic
 solution of equation (\ref{Nours}) into two branches is given by:
 \eqb \label{gbr}
 \gamma_{\rm b}= \frac{1}{\frac{1}{\gammamax}+\frac{2\alpha_{0}}{5c\tilde \Gamma}\left(r^{5/2}-r_{0}^{5/2}\right)}
 \eqe
 Taking into account $\gammamax\gg 1$ and $r \gg r_{0}$ equation
 (\ref{gbr}) becomes:
 \eqb
 \gamma_{\rm b}\approx \frac{5c\tilde
 \Gamma}{2\alpha_{0}}r^{-5/2}\equiv \gamma_{\rm c}
 \eqe
 Therefore, in this case the two characteristic Lorentz factors of
 the distribution $\gamma_{\rm c}, \gamma_{\rm b}$ coincide.
In order to obtain the asymptotic solutions presented in section
(\ref{constant}) one follows the procedure presented in the previous
sections.
\bibliographystyle{aa}
\bibliography{grb}

\end{document}